UDC 550.388.2

# Total electron content disturbances prior to Great Tohoku March 11, 2011 and October 23, 2011 Turkey Van earthquakes and their physical interpretation

## Zolotov O.V.[1], Namgaladze A.A.[1], Prokhorov B.E.[2,3]

[1] *Polytechnic Faculty of MSTU, Physics Department*
[2] *Helmholtz Centre Potsdam, GFZ German Research Centre for Geosciences*
[3] *University Potsdam, Applied Mathematics, Interdisciplinary Center for Dynamics of Complex Systems (DYCOS)*

**Abstract.** This paper discusses the GPS (Global Positioning System) observed TEC (Total Electron Content) variations prior to the M 9.0 Great Tohoku (Japan, Sendai) March 11, 2011 and M 7.1 Oct. 23, 2011 Turkey Van earthquakes as possible seismo-ionosphere precursors. We have formulated a set of the TEC phenomenological features often reported as precursors to strong earthquakes based on our experience and publications' analysis. This feature-set has been applied to the relative TEC deviations for time intervals March 08-11, 2011 and Oct. 20-23, 2011 preceding M 9.0 Great Tohoku (Japan) March 11 and Turkey Van Oct. 23 earthquakes, respectively. In both cases there have been revealed strong local long-living (of about several hours) TEC disturbances at the near-epicenter and magnetically conjugated areas. These disturbances may be treated as seismo-ionospheric precursors. The physical mechanism for the observed TEC structures for these two as well as for other cases of recent strong seismic events has been proposed. The anomalies have been interpreted and explained on the base of this physical mechanism from the origin hypothesis point of view in terms of electromagnetic lithosphere-ionosphere coupling.

**Аннотация.** В статье анализируются морфологические особенности вариаций полного электронного содержания (ПЭС) ионосферы перед землетрясениями 11 марта 2011 г. (Япония) и 23 октября 2011 г. (Турция) в качестве их возможных предвестников. На основе анализа существующих работ, а также собственных результатов исследований сейсмо-ионосферных предвестников землетрясений, сформулирован перечень наиболее часто называемых особенностей в ПЭС ионосферы как вероятных предвестников. Показано, что для рассматриваемых землетрясений в периоды 08-11 марта и 20-23 октября наблюдались аномальные возмущения ПЭС как долгоживущие структуры в околоэпицентральной и магнитосопряжённой к ней областях. Выявленные аномальные области обсуждаются с точки зрения возможного физического механизма их формирования – электромагнитной связи системы "литосфера-ионосфера".

**Ключевые слова:** полное электронное содержание, землетрясения 11 марта 2011 г. (Япония) и 23 октября 2011 г. (Турция), ионосферные предвестники землетрясений, ионосфера

## 1. Introduction

The lack of success in making forecasts of strong earthquakes triggered a series of interdisciplinary studies aimed at searching for the precursors in different physical parameters including seismo-ionosphere features' explorations (e.g., *Moore*, 1964; *Davies, Baker*, 1965; *Gokhberg et al.*, 1982; *Larkina*, 1989; *Parrot, Mogilevsky*, 1989; *Buchachenko et al.*, 1996; *Molchanov*, 1998; *Hayakawa et al.*, 2004; *Liu et al.*, 2006). Since middle-ninetieth the ionospheric TEC (Total Electron Content) became one of the most often used parameter analyzed for the possible manifestations of seismo-ionosphere coupling processes, see, e.g., (*Pulinets, Boyarchuk*, 2004) and references there in (*Afraimovich et al.*, 2004; *Zakharenkova et al.*, 2007; *Liu et al.*, 2004; 2011). The earthquakes that attract attention of the researchers are usually of large magnitude and great impact on the humans' environment.

This article is devoted to the main TEC phenomenological features' description, analysis and physical interpretation prior to the (a) near the East Coast of Honshu (Great Tohoku), Japan[1], M 9.0, March 11, 2011

---

[1] USGS catalog description at URL: http://earthquake.usgs.gov/earthquakes/eqinthenews/2011/usc0001xgp/

05:46 UT / 14:46 LT, (38.297°N, 142.372°E), D 30 km, and (b) Eastern Turkey[2] M 7.1, Oct. 23, 2011, 10:41 UT / 13:41 LT, (38.691°N, 43.497°E), D 16 km earthquakes.

To carry out the investigation systematically, we take into consideration the following aspects: (1) firstly, we define the signatures (i.e. the stable TEC features) as precursors or the TEC "anomalies" according to the publications available and our experience to look them for; (2) then we describe the revealed anomalies (their spatial and temporal features) for the case studies (3) and discuss an agreement and discrepancies between the anomalies and defined during step (1) pre-EQs TEC precursors' signatures from the physical mechanism of the pre-earthquake TEC disturbances generation point of view. We also take into account the possible influence of the solar and geomagnetic activity-driven disturbances on co-seismic TEC variations.

**2. TEC pre-earthquake disturbances' features**

Many investigators highlight the following features of the TEC disturbances before strong earthquakes as precursors to strong seismic events, see, e.g., (*Liu et al.*, 2006; *Pulinets*, 1998; *Pulinets, Boyarchuk*, 2004; *Ruzhin et al.*, 2002; *Zakharenkova et al.*, 2006; 2007). They are very often look like (1) local long-living TEC increase or depression regions that are situated in the vicinity of the earthquake epicenter area and often at the magnetically conjugated region. These anomalies do not propagate along the magnetic meridians in contrast to the ionospheric disturbances related to magnetic activity. The amplitude of plasma modification reaches the values of >30-90 % over the non-disturbed level. TEC positive modifications usually dominate (more often reported) according to *Zakharenkova et al.* (2006; 2007). (2) The vertical projection of the epicenter position does not mandatorily coincide with the maximum phenomenon's manifestation location. (3) The size of the anomaly maximum manifestation region depends on the earthquake's magnitude, extends larger than 1500 km in latitude and 3500-4000 km in longitude, i.e. agrees with the *Dobrovolsky et al.* (1979) estimation. The shapes and dimensions of the disturbed areas are kept rather stable during 4-8 hours appearing a few days before the seismic event. (4) The anomalies are reported from several days or hours to couple of weeks before the earthquake release moment. (5) In case of the strong low-latitudinal earthquakes there are effects related to the modification of the ionospheric F2-region equatorial anomaly: increase or decrease of the equatorial anomaly with the trough deepening or filling (*Depueva, Ruzhin*, 1995; *Pulinets et al.*, 2003; *Depueva et al.*, 2007).

Based on the TEC relative (%) variations' analysis before a few strong recent seismic events (*Namgaladze et al.*, 2011a,b; 2012; *Namgaladze, Zolotov*, 2011; 2012; *Zolotov et al.*, 2011; 2012) we also extend the named above set of pre-earthquake TEC features with terminator and sub-solar point effects: (i) the "ban"-time takes place for the anomalies to exist corresponding to the subsolar point being at the near-epicenter area, i.e. near-noon hours. At this time anomalous TEC relative deviations are reduced in general down to almost full destruction. (ii) There is a link between the anomalous regions' shape and position and terminator: in general, with the sunrise terminator coming we see depression of the anomalies and their partial shift from the terminator towards the night-sector. After the sunset terminator leave we see the anomalies renewal.

**3. TEC disturbances' case studies**
**3.1. TEC disturbances' method**

To estimate pre-earthquake TEC variations we have built differential TEC maps relative to the quiet background conditions. We used global ionospheric maps (GIM) of the TEC (*Dow et al.*, 2009) provided by the NASA in IONEX format as the initial data for the analysis, namely the igsg product. The spatial resolution of those TEC data is 5° in longitude and 2.5° in latitude (geographic longitude vs latitude). Corresponding time resolution is 2 hours, the accuracy is 2-8-TECU, where 1 TECU = $1 \cdot 10^{16}$ el/m$^2$. In the present investigation we defined and calculated the background TEC values (i.e. undisturbed conditions) as 7-days UT-grouped running observations' medians before the current calculation moment.

**3.2. Case study of the TEC variations before M 9.0 Great Tohoku March 11, 2011 earthquake**

The geomagnetic activity during considered period was quiet-to-moderate with episodic periods of moderate-to-strong disturbances. On March 6-10, 2011 the Dst index was within the interval of from -22 nT to 4 nT without any significant disturbances. On March 10 a moderate magnetic storm took place: Dst index varied from -7 nT on March 10 down to -82 nT on March 11. The storm sudden commencement took place on March 10. The Kp index during March 6-9 in general did not exceed 3.

According to the USGS Significant Earthquake and News Headlines Archive, the investigation of the considered period was complicated by a series of strong (M>5) seismic events happened during March 7-11, 2011:

(a) March 07, 2011, 00:09:38 UT, M6.6 – Solomon Islands[3], (10.334°S, 160.739°E), D37.9 km;

---

[2] USGS catalog description at URL: http://earthquake.usgs.gov/earthquakes/eqinthenews/2011/usb0006bqc/

(b) March 09, 2011, 02:45:20 UT, M7.3 – near the East coast of Honshu, Japan[4], (38.440°N, 142.840°E), D32 km;

(c) March 09, 2011, 21:24:51 UT, M6.5 – New Britain Region, Papua New Guinea[5], (6.022°S, 149.659°E), D29 km;

(d) March 10, 2011, 04:58:13 UT, M5.5 – Myanmar-China Border Region[6], (24.727°N, 97.597°E), D10 km;

(e) March 11, 2011, 05:46:24 UT, M9.0 – near the East coast of Honshu, Japan[7], (38.322°N, 142.369°E), D30 km.

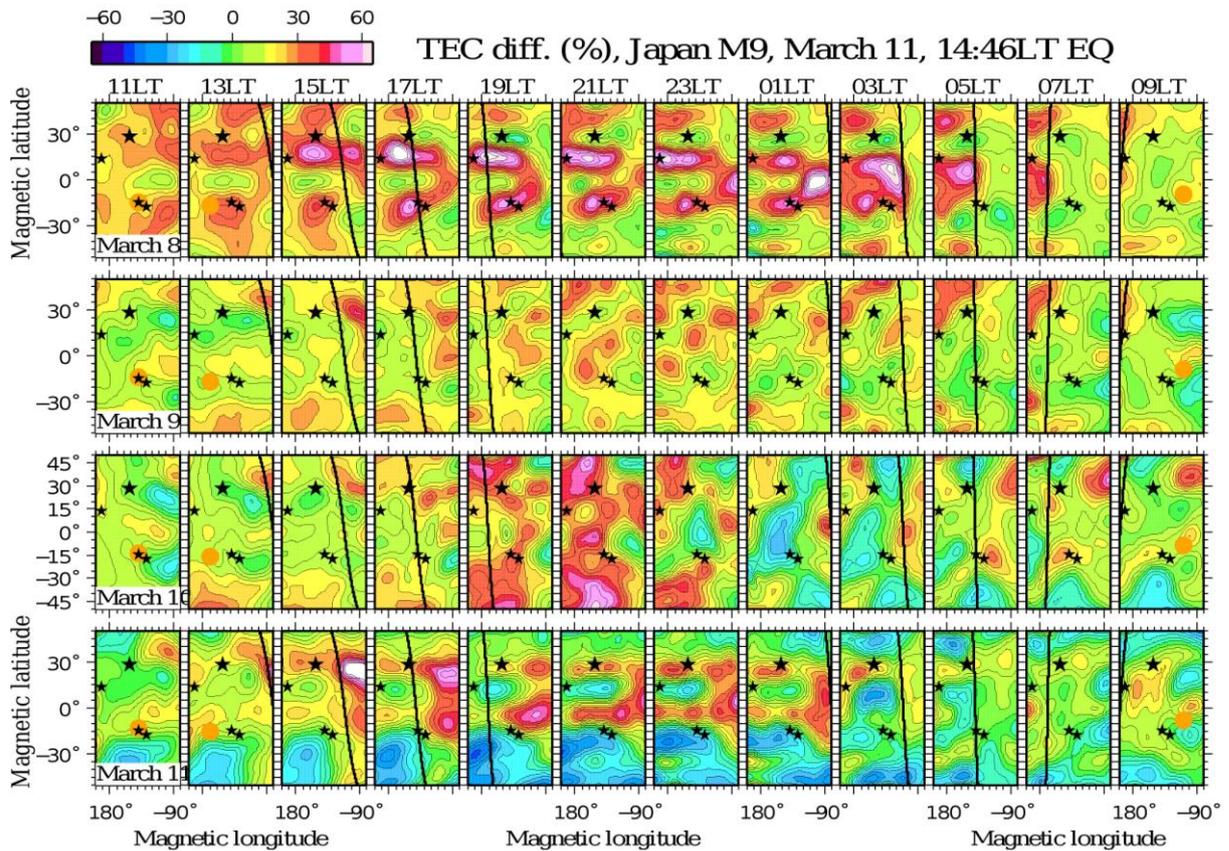

Fig. 1. The TEC deviations (%) from the background values for 02UT/11LT-24UT/09L (from left to right) March 8-11 (from top to bottom), 2011. Star denotes the earthquake epicenter position. Black curve – terminator. Orange circle – the subsolar point position

As one can see, the epicenters' of at least more than two earthquakes are within the region of 2500 km vs 4500 km which is the typical reported size of the pre-earthquake TEC anomaly manifestation area. The time-lag between the considered events is less than a week, i.e. less than the reported time-interval (7 or event 15 days and more according to some authors) of the seismo-associated TEC disturbances. Therefore, those criteria do not allow us to separate each earthquake's deposit into the ionosphere TEC disturbances, and we consider anomalous TEC modification as the resulting from action of a set of seismic sources (each with its own acting regime). Basing on the epicenters locations' spatial distribution and keeping in mind that TEC anomalies treated as precursors are reported at the near-epicenter area, one may assume that equatorial earthquakes make major deposit into Appleton anomaly seismo-modifications, mid-latitudinal – into the mid-latitudinal TEC disturbances. But this topic is not evident and the latitudinal spread of TEC effects from the source is a topic for on-going research. At least, modifications of the Appleton (equatorial ionization) anomaly take place in form of crests shifts and "trough" modifications, but it (the anomaly) also should react on electric fields penetrating towards equator from mid-latitudes.

---

[3] USGS catalog description at URL: http://earthquake.usgs.gov/earthquakes/eqinthenews/2011/usb0001q9i/
[4] USGS catalog description at URL: http://earthquake.usgs.gov/earthquakes/eqinthenews/2011/usb0001r57/
[5] USGS catalog description at URL: http://earthquake.usgs.gov/earthquakes/eqinthenews/2011/usc0001wfq/
[6] USGS catalog description at URL: http://earthquake.usgs.gov/earthquakes/eqinthenews/2011/usc0001wnu/
[7] USGS catalog description at URL: http://earthquake.usgs.gov/earthquakes/eqinthenews/2011/usc0001xgp/

To reveal main phenomenological features of the TEC relative disturbances before the M 9.0 Great Tohoku (Japan) March 11 earthquake we analyzed TEC disturbances (%) maps, presented in Fig. 1 for March 08-11, 2011. As one can see, on March 8, 2011 anomalies took place as positive structures along the parallel and situated at both sides of the geomagnetic equator. These anomalies existed during 04 UT / 13 LT – 20 UT / 05 LT, spatially occupied up to ~20˚×~25˚ (latitude × longitude) reaching values from >40 % up to >60 % in Northern hemisphere; and up to ~15˚×25˚ (latitude × longitude) reaching magnitudes from ~30 % up to >50 %; they firstly were formed at the near-epicenter areas, then their magnitude increased. During 08 UT / 17 LT – 12 UT / 21 LT the anomalies spread out along the parallel at both hemispheres, occupying a larger region. There is a strong tendency to fulfill the Appleton's anomaly from 12 UT / 21 LT, and from 18 UT / 03 LT one can see a completely formed unite positive structure occupying ~30˚×25˚ (magnetic longitude × magnetic latitude) and up to 60 % by magnitude. The TEC anomaly comprised an ellipse-like region with the earthquake round its borders. One also can see that disturbances are stronger in the Northern hemisphere, the terminator and subsolar point income trigger the reduction of the anomaly up to its' almost full destruction. The forms of the anomaly are kept rather stable, the magnitude initially grows, then stays constant or slightly decreases until the sunset terminator come.

The passages of the sunrise terminator and later of the subsolar point degraded the TEC anomaly and led to its almost full decay in the case of March 09, i.e. on the day of the M 7.2 March 09, 02:45 UT / 11:45 LT earthquake (happened at nearly the same exact positions of the M9 March 11) and the M6.5, (6.022°S, 149.659°E), 21:24 UT / 06:24 LT earthquakes.

On March 10, 2011 during 10-14 UT / 19-23 LT some strong positive disturbances that moved from high-to-low latitudes existed. We attributed them to the geomagnetic activity.

### 3.3. Case study of the TEC variations before Turkey Van Oct. 23, 2011 earthquake

Oct. 10-23 time-interval was rather quiet: Dst index was within ±20 nT during whole period. A geomagnetic storm took place on the midnight between Oct. 24 & 25, i.e. after the considered Turkey Van EQ happened and, therefore, could not mask the TEC variations earlier. Kp index was in general less then 3 during Oct. 10-23 and less then 1 during Oct. 22-23. The storm sudden commencement happened on Oct. 24, i.e. out of the considered interval.

During Oct. 20-23, 2011 we see effects (see Fig. 2) associated to seismic activity at the near-epicenter, magnetically conjugated to it and equatorial regions. On Oct. 20 positive TEC structures existed from 14 UT /17 LT to 20 UT / 23 LT at the near-epicenter area and occupied ~10˚ along the meridian and ~15˚ along the parallel, reaching ~30 % by magnitude. The TEC increase areas were shifted toward equator and to the west. At the magnetically conjugated region the TEC increases up to >30 % existed from 14 UT /17 LT to 20-22 UT / 23-01 LT on the area of about 25˚ along the meridian and ~40˚ along the parallel. They also were shifted towards equatorial region. At the equator the negative structures of about -15 % were observed during 10-20 UT / 13-23 LT. From 20 UT / 23 LT till 24 UT / 01 LT the positive and negative structures formed a single positive spot-like disturbed area that degraded with the terminator income. During near-noon hours (04-12 UT /07-15 LT) no stable long-living disturbances were observed.

On Oct. 21 the disturbed TEC structures manifested more evidently. The TEC enhancements reached >40 % by magnitude, existed from 14 UT / 17 LT to 20 UT / 23 LT and occupied ~15˚ via latitude and ~25˚ via longitude. These increases were also shifted equatorwards and westwards. Positive structures at the magnetically conjugated area reached up to >45 % values by magnitude during 14-22 UT and occupied over 10-15˚ via latitude and ~35˚ via latitude. Round the magnetic equator the TEC reductions ~-15 % happened during 14-20 UT. The terminator approaching also was followed by a "single spot"-like structure formation with consequent anomaly reduction during 22-24 UT. During near-noon hours no significant anomalies were observed.

On Oct. 22 during 14-20 UT the TEC enhancements were in general of about 30 % reaching maximum value up to >35 % at 18 UT on a region of about 10-20˚ via longitude and latitude. At the same period the TEC enhancements of about 45 % existed at the magnetically conjugated area also reaching maximum at 18 UT. We also see some negative structures near the magnetic equator. All anomalous regions as in previous days transformed into a single spot structure and then degraded.

Oct. 23 is the day of EQ, which happened at 10:41 UT, i.e. within the near-noon time interval; no significant anomalies were observed before the earthquake at the day of the earthquake. Starting from 14-16 UT some positive disturbances existed and they looked like "blurring" of the previous day disturbances. We expect they might be post-effects of the considered earthquake.

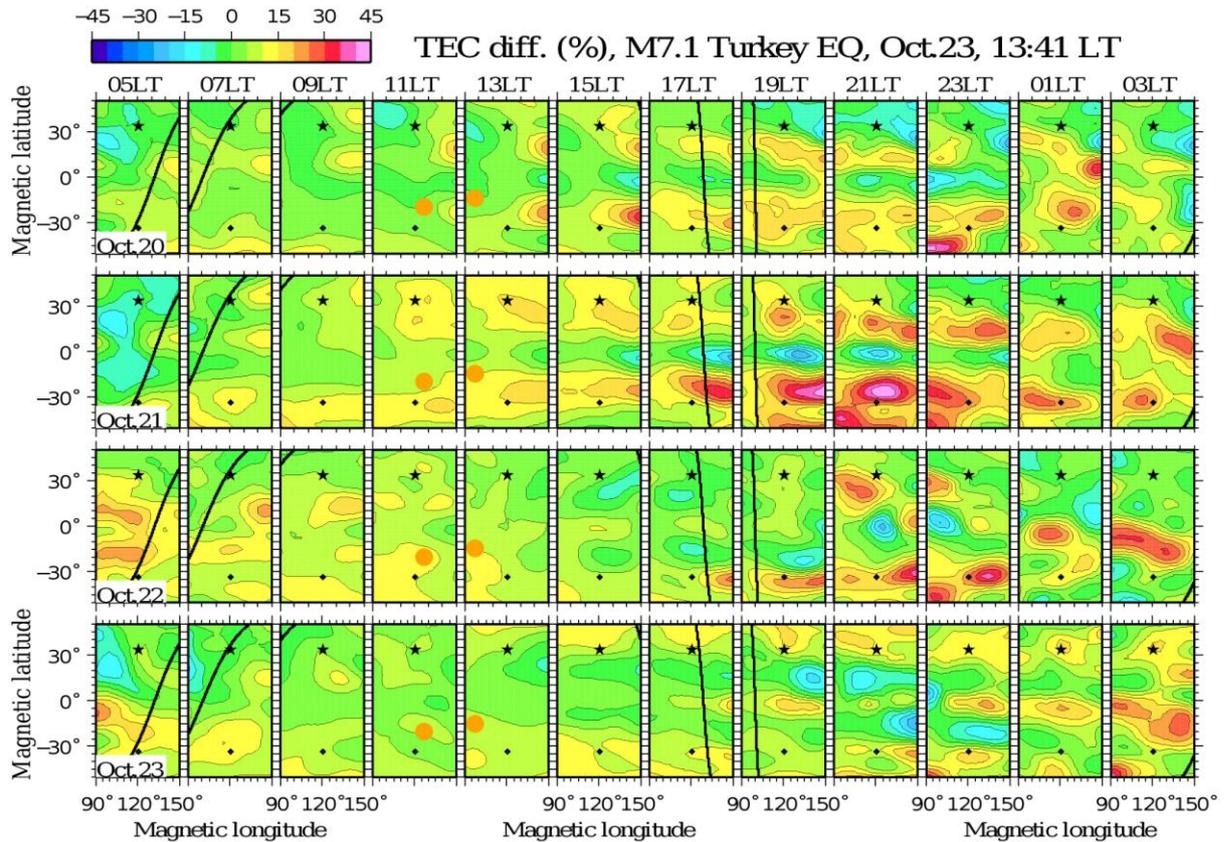

Fig. 2. The TEC deviations (%) from the background values for 02UT/05LT-24UT/03L (from left to right) Oct. 20-23 (from top to bottom), 2011. Star denotes the earthquake epicenter position. Diamond – the magnetically conjugated point. Black curve – terminator. Orange circle – the subsolar point position

At the area of the epicenter and magnetically conjugated points we had two types of stable structures: (1) positive structures at both the epicenter and magnetically conjugated areas followed ("overshielded") by negative ones and (2) negative TEC disturbances at the near-epicenter area with corresponding TEC enhancements at the magnetically conjugated region both followed by the opposite sign disturbances round or near its' borders.

The most evident positive TEC structures manifestation at both hemispheres happened on Oct. 21, 2011. It reached values up to >45 % by magnitude, occupied spatial region ~15º lon. vs ~20º lat. near the epicenter and ~25-35º lon. vs 15-20º lat. at the magnetically conjugated area. The lifetime of TEC enhancements was 4-6 hours or more.

**4. Discussion**
**4.1. Great Tohoku (Japan) March 11, 2011 and Turkey Van Oct. 23, 2011 case studies vs. pre-Eqs' TEC feature-set & other cases**

From the relative (%) TEC disturbances maps description and analysis provided above one may mark out the following common pre-earthquake TEC variations features in both considered cases: (1) TEC anomalies were observed a few days before the main shock event, linked to the near-epicenter and magnetically conjugated regions, also followed by Appleton anomaly's modifications, and did not propagate along the meridians. (2) Their lifetime was in general limited to approximately 15LT-04LT (varied from >4 hours to 8 hours, cf. Fig. 1 vs Fig. 2), reaching disturbances magnitude maximum up to >40-60 %, occupying spatial area of about 10-25º via latitude and 15-35º or more via longitude. (3) TEC anomalous structures were rather stable.

This behavior corresponds to the presented in section 2 of this article pre-earthquake TEC feature-set. Additionally one may see the effects related to the terminator and subsolar point position, i.e. solar-illumination driven effect: the income of sunrise terminator initiate TEC disturbanses' distortion with up to almost full reduction under the near-subsolar point region. The sunset terminator leaving is followed by the anomalies renewal. Such behavior exactly matches the additional features reported in section 2 of this paper as well as in cases of strong Haiti, M7.0, Jan. 12, 2010, and Argentina, M7.0, Jan. 01, 2011 and Chile, M7.1, Jan. 02, 2011 earthquakes (cf. Figs. 1, 2, 3). One also may see that in both presented cases TEC anomalies for about 4 hours before the terminator arrival are shifted equatorwards with consequent a single-spot structure formation and its following degradation.

We also have ability to compare our results for the M 9.0 Great Tohoku (Japan) EQ with *Xu et al.* (2011) investigation, whose aim was declared as to report the existence of the related changes at the Earth surface and in the ionosphere, and show the potential ability of the application of multi-source data to identify seismic precursors. In short, they reported: (1) Significant TEC enhancements were detected. The maximum deviation was of about ~+30 TECu. It started at 06 UT of March 08, 2011. (2) They were localized in 100-160°E longitudinal sector, in the area of Northern Appleton Anomaly crest. Similar effects were also observed at the magnetically conjugated region. (3) The most long-living TEC disturbances lasted ~20 hours. Moving disturbances were reported on March 7 & 9, 2011 and were attributed to the solar radiation. *Xu et al.* (2011a) also noted shift of the seismo-associated anomalies from the earthquake epicenter. As we see, our and *Xu et al.* (2011a) results are in agreement. A few differences exist: they speak nothing about terminator and subsolar point-related effects; some minor differences in TEC magnitudes and lifetime. We expect this is due to the reason that *Xu et al.* (2011a) analyzed TEC deviations in absolute units (namely, TECu) and used different sliding window size (27 days) to calculate the TEC quiet background variation.

We expect the named above features are principal for understanding the nature of the phenomena and it will be discussed in the next subsection.

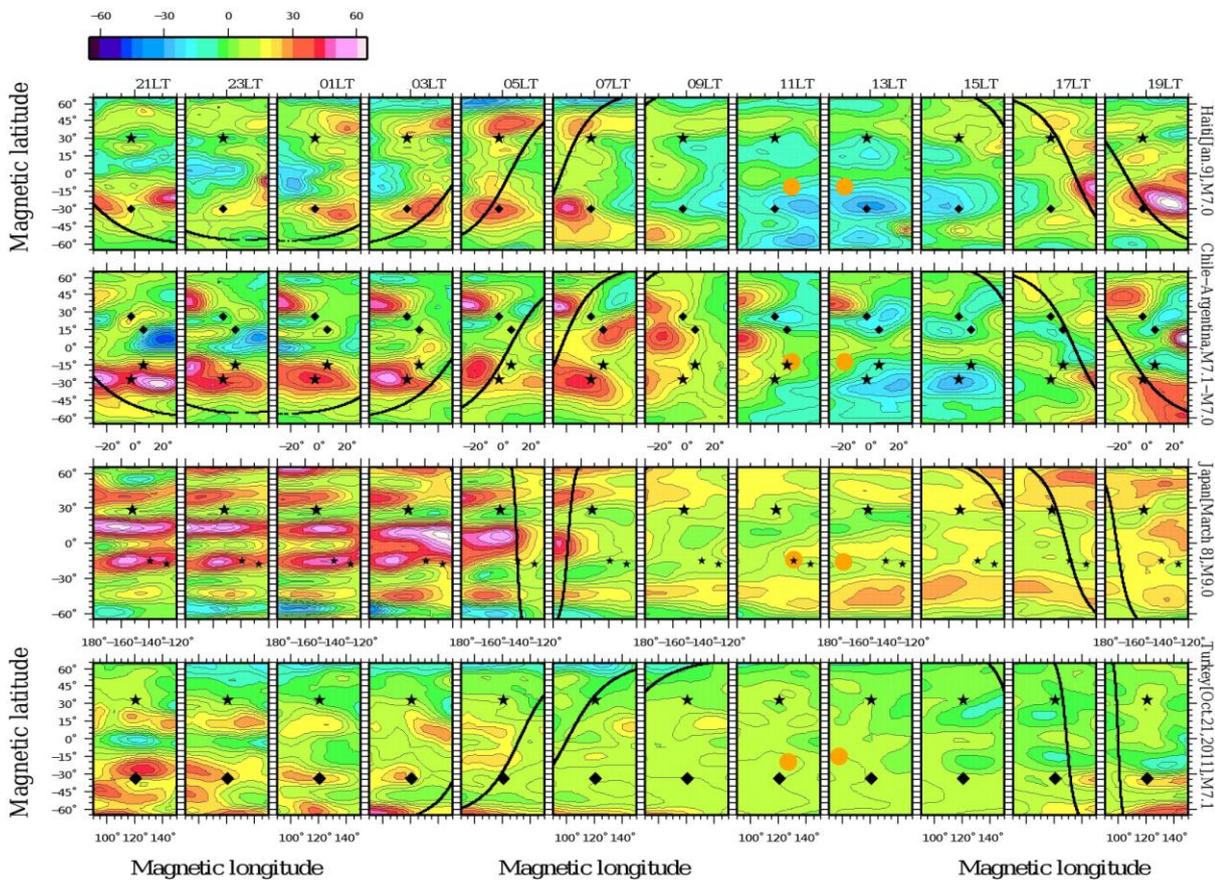

Fig. 3. The TEC deviations (%) from the background values for 21LT-19LT before (from top to bottom) (1) Haiti Jan. 12, 2010 EQ; (2) Chile & Argentina Jan, 1-2, 2011 Eqs; (3) Great Tohoku (Japan) March 11, 2011 EQ; (4) Turkey Van Oct. 23, 2011 EQ. Orange circle – the subsolar point. Black curve – terminator. Star – the epicenter position. Diamond – magnetically conjugated point

**4.2. Morphological features' interpretation: physical mechanism and origin**

There are a few channels of penetration of the earthquake preparation processes' impact through the underlying neutral atmosphere into ionosphere: (1) wave channel including AGW; (2) electromagnetic channel, etc. A detailed discussion of various mechanisms is out of this investigation and may be found in, e.g., (*Pulinets, Boyarchuk*, 2004; *Liperovsky et al.*, 2008; *Ondoh*, 2009; *Uyeda et al.*, 2009; *Hayakawa, Hobara*, 2010).

We reject the neutral atmosphere wave channel due to absence of the wave signatures of the discussed TEC variations and strongly relay on the electromagnetic mechanism of the lithosphere-atmosphere-ionosphere coupling. *Namgaladze et al.* (2007; 2009a,b) consider that the most probable reason of the NmF2 and TEC disturbances observed before the earthquakes is the vertical drift of the F2-region ionospheric plasma under the influence of the zonal electric field of seismogenic origin. In the middle latitudes the upward electromagnetic drift, created by the eastward electric field, leads to the increase of the NmF2 and TEC due to the plasma transport to the regions with lower concentration of the neutral molecules and, consequently, with lower loss rate of dominating ions O+ in the ion-molecular reactions. The electric field of the opposite direction (westward) creates the opposite – negative – effect in NmF2 and TEC. In the low latitude regions (near the geomagnetic equator) the increase of the eastward electric field leads to the deepening of the Appleton anomaly minimum ("trough" over the magnetic equator in the latitudinal distribution of electron concentration) due to the intensification of the fountain-effect.

It should be noted that the seismo-generated zonal electric field does not act alone and interact with regular drift pattern, thus, changing it. The meridional electric field also influences on the electron density variations. Therefore 3D consideration is required. Numerical simulations (*Namgaladze et al.*, 2007; 2009a,b) using 3D time-dependent Upper Atmosphere Model (*Namgaladze et al.*, 1988; 1991; 1998a,b) showed that additional eastward electric fields of about ~1-4 mV/m in case of the low-latitudes and ~4-10 mV/m in case of the mid-latitudes are required to produce the observed TEC disturbances. They exceed the background quiet fields (of about 0.2 and 1 mV/m, correspondingly), but they are noticeably smaller than the quiet high-latitude electric fields of magnetospheric origin (15-25 mV/m) obtained in the model calculations. Model simulated seismogenic electric fields' magnitudes agree with the INTERCOSMOS-BULGARIA-1300 satellite observations (*Chmyrev et al.*, 1989; *Gousheva et al.*, 2006; 2008a,b; 2009), rocket measurements of large intense electric fields in the E layer of ionosphere (*Yokoyama et al.*, 2002) and ionosonde data-derived electric fields' estimations (*Xu et al.*, 2011b) as well as with recent other authors' simulations, e.g., (*Klimenko et al.*, 2011; 2012; *Liu et al.*, 2011).

This mechanism requires an explanation the way such additional electric fields appear at the ionospheric heights. Many authors (e.g., *Pulinets*, 1998; *Pulinets, Boyarchuk*, 2004; *Sorokin et al.*, 2005; 2006; 2007) associate this hypothetical seismogenic electric field with the vertical turbulent transportation of the injected aerosols and radioactive particles (radon isotopes). The increase of the atmospheric radioactivity level during the earthquake preparation leads to the changes of the ionization and electric conductivity of the near-ground atmosphere. The joint action of these processes leads to the generation of the electric field in the ionosphere up to the value of units-tens mV/m (*Chmyrev et al.*, 1989). *Sorokin et al.* (2005; 2006; 2007) have calculated the ionospheric electric field related with external electric current variations in the lower atmosphere. This current is formed due to the convective upward transport of charged aerosols and their gravitational sedimentation in the lower atmosphere. This effect is related with the occurrence of ionization source due to seismic-related emanation of radon and other radioactive elements into the lower atmosphere.

Thus, the principle idea is local (regional) change of the resistance of the neural atmosphere column over the epicenter area and corresponding appearance of the vertical electric current. This scheme may be generalized by assuming not only radon and radioactive gases exhalation, but other sources of near-ground atmosphere ionization. *Freund et al.* (2009), *Freund* (2011) proposed another mechanism of the near-ground atmosphere layer ionization based on the so-called "positive holes": most crustal rocks contain dormant electronic charge carriers in the form of peroxy defects; when rocks are stressed, peroxy links break, releasing electronic charge carriers, known as positive holes. The positive holes are highly mobile and can flow out of the stressed subvolume. F. Freund expects this mechanism to be significantly more efficient than the above-named radon-related ones. It also should be noticed that mentioned sources of near-ground air ionization and vertical electric current formation mechanisms do not prohibit each other and common action is possible.

Recent UAM simulations' results (*Namgaladze*, 2010; *Namgaladze et al.*, 2011a,b; *Namgaladze, Zolotov*, 2011) show that a vertical electric current density of about $10^{-8}$ A/m$^2$ at an area of about 1000 km × 4000 km can create the electric fields of several mV/m in the nighttime ionosphere which in turn can produce the TEC variations of up to 50 %, which turns out to be very similar to the observed ones. Those model results are in considerable agreement with (1) *Sorokin et al.* (2005; 2006; 2007) simulations, that used the external current density of about $10^{-6}$ A/m$^2$ at the area of about 200 km in radius (approximately 130 000 km$^2$) to create the electric field of about several mV/m in the ionosphere; (2) as well as with *Kuo et al.* (2011) results who used current density 0.01-1 μA/m$^2$ over area 200 km × 30 km to produce 1-30 % TEC night-time variations. Note they used stronger current density but set on a smaller spatial region.

It should be noted that the above current density magnitudes are significantly larger than "fair-weather" vertical current between the Earth and the ionosphere. If electric current is assumed as $j = nq\mathbf{v}$, where $n$ is the concentration of charged particles, $q$ is the charge and $\mathbf{v}$ is the velocity vector, then the model external vertical electric current density value (~$10^{-8}$A/m$^2$, required for >50 % TEC disturbances) is about $10^4$ times of average vertical electric current density. Therefore, provided that **v** does not drastically changes, we can estimate $n_d \sim 10^4 n_Q$, where $n_d$ is the seismo-disturbed concentration and $n_Q$ corresponds to the quiet condition (with usual current density of 2-3 pA/m$^2$).

Nevertheless, the required vertical electric current magnitudes are significantly less than *Sorokin et al.* (2005; 2006; 2007) and *Kuo et al.* (2011) estimations as well as significantly less than *Freund* (2011) magnitudes obtained in heavy load experiment (values up to 10-100 A/km$^2$). Moreover, our used in simulations current densities (needed to generate >50 % TEC disturbances) are close to the upper limits of the measured vertical currents over thunderstorm regions and tropical forests (*Making, Ogawa*, 1984; *Davydenko et al.*, 2009; *Le Mouël et al.*, 2010), but do not exceed them.

It is evident that such intensive currents are significant part of the Global Electric Current (*Mareev*, 2010) and unable to exist long time and correspond to very extreme situations of strong earthquakes preparation accompanied by the appearance of powerful ionization sources over tectonic faults and large ionospheric TEC disturbances. Regular observations of those currents are hard to realize due to the absence of dense network of monitoring stations which measure the vertical currents itself but not electric fields (and require the knowledge of the conductivity) or magnetic disturbances (affected by ionospheric and magnetospheric noise).

**4.3. TEC anomalies from the electromagnetic coupling point of view**

Taking into account the electromagnetic lithosphere-ionosphere coupling hypothesis we make an effort to physically interpret the origin of essential TEC features reported.

*Magnetic conjugation of the reported anomalies*

The electric fields, if appear at one hemisphere, easily propagate into the opposite one. The geomagnetic field lines are "ideal" conductors with "infinite" conductivity and so the electric field is "simultaneously" transferred to the other basement of the flux tube, i.e. at the magnetically conjugated region. That will lead to the same electric potential and electric fields at both hemispheres but the effects of such fields do not mandatory the same and may differ due to the differences in the neutral atmosphere and ionosphere states. Both case studies report geomagnetic conjugation of the observed phenomena.

*Terminator and subsolar point-related effects*

The electric fields are highly dependent on the local conductivity: (1) the income of the well-conducting sunlit ionosphere corresponds to the income of the enhanced conductivity areas, (2) which in turn should significantly decrease electric field up to its full disappearance (3) and corresponding TEC disturbances removal possibly with time-lag due to the ionospheres' inertness. Exactly the same time-evolution of the TEC earthquakes forerunners we see here and as also reported in other case studies (*Namgaladze et al.*, 2011a,b; *Namgaladze, Zolotov*, 2011; *Zolotov et al.*, 2011), which is additional evidence in favor of the named above hypothesis.

*Appleton anomaly modifications*

Equatorial ionization (Appleton) anomaly is strongly driven by the electric fields, therefore, the appearance of additional electric field of seismic origin should lead to the Appleton anomaly modification as a unite structure.

*The linkage of the anomalies to the fixed geo-position*

It corresponds to the fixed position of disturbance source if it is related to the fault.

*Life time of the anomalous phenomena*

The life time of the anomaly time of appearance may be explained if the stress process is going long enough and TEC disturbances are generated taking into account named above terminator and subsolar point effects.

**5. Conclusions**

This paper presents a new extended set of pre-earthquake TEC disturbances' features and phenomenological description of the TEC precursors to earthquakes. To validate the approach, it was applied to investigate two strong recent earthquakes.

It was shown that in case of Great Tohoku (Japan) M9.0 EQ the most pronounced anomalies happened on March 8, 2011 during 04 UT – 20 UT. They looked as the TEC enhancements of about 40-60 % along the geomagnetic parallel on both sides of the geomagnetic equator. The TEC anomalies comprised two ellipse-like regions of about ~20°×~25° (latitude × longitude) near the EQ epicenter and the magnetically conjugated area. The passages of the sunrise terminator and later of the subsolar point degraded the anomaly and led to its decay in the case of the whole day March 09.

In case of Turkey Van M7.1 EQ, TEC enhancements (see http://goo.gl/zRW9H) of about ~40 % were observed during Oct. 20-23, 2011 both near the epicenter and magnetically conjugated areas and reached maximum on Oct. 21. They existed from 14 UT to 20 UT and occupied ~10-15° × ~25-35° size areas. The terminator approaching was preceded by single spot structure formation with consequent anomaly reduction during 22-24 UT. TEC pre-EQs disturbances' magnitudes for both cases are relative to the quiet variation derived as 7 days UT-grouped running medians.

In both cases major TEC features persisted: (1) local long-living TEC increases situated near the earthquake near-epicenter and magnetically conjugated areas. These anomalies do not propagate along the meridians. The amplitude of plasma modification reaches the values of >40-60 %. (2) The vertical projection of the epicenter position does not coincide with the maximum phenomenon's manifestation location. The shapes and dimensions of the disturbed areas are kept rather stable. (3) There are effects related to the modification of the ionospheric F2-region equatorial anomaly. (4) There are strong subsolar point and terminator-related effects.

The features revealed for the Turkey and Japan EQs were compared with two other strong recent EQs and with the TEC seismo-precursors' features-set. We also compared Japan EQ results with *Xu et al.* (2011) investigation. A satisfactorily agreement was revealed.

All special features like magnetic conjugation of the observed phenomena, terminator and subsolar point-related effects, Appleton's anomaly modification, linkage to geoposition and the lifetime of the anomalies were discussed and explained in terms of electro-magnetic physical mechanism of lithosphere-ionosphere coupling based on the F2-layer plasma [**E**×**B**] drift under influence of seismogenic electric fields.

**Acknowledgements.** The authors acknowledge (1) NOAA National Geophysical Data Center / NGDC for the Ap, Kp and Dst indexes' data (http://spidr.ngdc.noaa.gov/spidr/); (2) U.S. Geological Survey and National Earthquake Information Center for the earthquakes data, URL: http://earthquake.usgs.gov; (3) IGS community and Crustal Dynamics Data Information System (CDDIS) for the TEC Global Ionosphere Maps (*Noll*, 2010); (4) University of Hawaii and GMT community for General Mapping Tools (*Wessel, Smith*, 1998).

**References**

**Afraimovich E.L., Astafieva E.I., Gokhberg M.B., Lapshin V.M., Permyakova V.E., Steblov G.M., Shalimov S.L.** Variations of the total electron content in the ionosphere from GPS data recorded during the Hector Mine earthquake of October 16, 1999, California. *Russian Journal of Earth Sciences*, v.6, N 5, p.339-354, 2004.

**Buchachenko A.L., Oraevskii V.N., Pokhotelov O.A., Sorokin V.M., Strakhov V.N., Chmyrev V.M.** Ionospheric precursors to earthquakes. *Phys. Usp.*, v. 39, p. 959-965, doi:10.1070/PU1996 v039n09ABEH001550, 1996.

**Chmyrev V.M., Isaev N.V., Bilichenko S.V., Stanev G.A.** Observation by space-borne detectors of electric fields and hydromagnetic waves in the ionosphere over on earthquake center. *Phys. Earth and Planet. Inter.*, v.57, p.110-114, doi:10.1016/0031-9201(89)90220-3, 1989.

**Davies K., Baker D.M.** Ionospheric effects observed around the time of the Alaskan earthquake of March 28, 1964. *J. Geophys. Res.*, v.70, N 9, p.2251-2253, doi:10.1029/JZ070i009p02251, 1965.

**Davydenko S.S., Thomas C.M., Maribeth S.** Modeling the electric structures of two thunderstorms and their contributions to the global circuit. *Atmos. Res.*, v. 91, N 2-4, p.165-177, 2009.

**Depueva A.Kh., Ruzhin Yu.Ya.** Seismoionospheric fountain-effect as analogue of active space experiment. *Adv. Space Res.*, v.15, N 12, p.151-154, 1995.

**Depueva A.Kh., Mikhailov A.V., Devi M., Barbara A.K.** Spatial and time variations in critical frequencies of the ionospheric F region above the zone of equatorial earthquake preparation. *Geomagnetism and Aeronomy*, v.47, N 1, p.129-133, doi:10.1134/S0016793207010197, 2007.

**Dobrovolsky I.R., Zubkov S.I., Myachkin V.I.** Estimation of the size of earthquake preparation zones. *Pure and Applied Geophysics*, v.117, N 5, p.1025-1044, doi:10.1007/BF00876083, 1979.

**Dow J.M., Neilan R.E., Rizos C.** The International GNSS Service in a changing landscape of Global Navigation Satellite Systems. *J. Geodesy*, v.83, N 3-4, p.191-198, doi:10.1007/s00190-008-0300-3, 2009.

**Freund F.T., Kulahci I.G., Cyr G., Ling J., Winnick M., Tregloan-Reed J., Freund M.M.** Air ionization at rock surfaces and pre-earthquake signals. *J. of Atmospheric and Solar-Terrestrial Physics*, v.71, p.1824-1834, 2009.

**Freund F.** Pre-earthquake signals: Underlying physical processes. *J. Asian Earth Sci.*, v.41, N 4-5, p.383-400, doi:10.1016/j.jseaes.2010.03.009, 2011.

**Gokhberg M.B., Morgounov V.A., Yoshino T., Tomizawa I.** Experimental Measurement of Electromagnetic Emissions Possibly Related to Earthquakes in Japan. *J. of Geophysical Research*, v.87(B9), p.7824-7828, doi:10.1029/JB087iB09p07824, 1982.

**Gousheva M., Danov D., Hristov P., Matova M.** Ionospheric quasi-static electric field anomalies during seismic activity in August-September 1981. *Nat. Haz. Earth Syst. Sci.*, v.9, N 1, p.3-15, doi:10.5194/nhess-9-3-2009, 2009.

**Gousheva M.N., Glavcheva R.P., Danov D.L., Hristov P.L., Kirov B.B., Georgieva K.Y.** Electric field and ion density anomalies in the mid latitude ionosphere: Possible connection with earthquakes? *Adv. Space Res.*, v.42, N 1, p.206-212, doi:10.1016/j.asr.2008.01.015, 2008a.

**Gousheva M., Danov D., Hristov P., Matova M.** Quasi-static electric fields phenomena in the ionosphere associated with pre- and post earthquake effects. *Nat. Haz. Earth Syst. Sci.*, v.8, N 1, p.101-107, doi:10.5194/nhess-8-101-2008, 2008b.

**Gousheva M., Glavcheva R., Danov D., Angelov P., Hristov P., Kirov B., Georgieva K.** Satellite monitoring of anomalous effects in the ionosphere probably related to strong earthquakes. *Adv. Space Res.*, v.37, N 4, p.660-665, doi:10.1016/j.asr.2004.12.050, 2006.

**Hayakawa M., Hobara Y.** Current status of seismo-electromagnetics for short-term earthquake prediction. *Geomatics, Natural Hazards and Risk*, v.1, N 2, p.115-155, doi:10.1080/19475705.2010.486933, 2010.

**Hayakawa M., Molchanov O.A., NASDA/UEC Team.** Summary report of NASDA's earthquake remote sensing frontier project. *Phys. Chem. Earth*, v.29, N 4-9, p.617-625, doi:10.1016/j.pce.2003.08.062, 2004.

**Klimenko M.V., Klimenko V.V., Zakharenkova I.E., Pulinets S.A., Zhao B., Tsidilina M.N.** Formation mechanism of great positive TEC disturbances prior to Wenchuan earthquake on May 12, 2008. *Adv. Space Res.*, v.48, N 3, p.488-499, doi:10.1016/j.asr.2011.03.040, 2011.

**Klimenko M.V., Klimenko V.V., Zakharenkova I.E., Pulinets S.A.** Variations of equatorial electrojet as possible seismo-ionospheric precursor at the occurrence of TEC anomalies before strong earthquake. *Adv. Space Res.*, v.49, N 3, p.509-517, doi:10.1016/j.asr.2011.10.017, 2012.

**Kuo C.L., Huba J.D., Joyce G., Lee L.C.** Ionosphere plasma bubbles and density variations induced by pre-earthquake rock currents and associated surface charges. *J. Geophys. Res.*, v.116, A10317, doi:10.1029/2011JA016628, 2011.

**Larkina V.** Some statistical results on very low frequency radiowave emissions in the upper ionosphere over earthquake zones. *Physics of the Earth and Planetary Interiors*, v.57, N 1-2, p.100-109, doi:10.1016/0031-9201(89)90219-7, 1989.

**Le Mouël J.-L., Gibert D., Poirier J.-P.** On transient electric potential variations in a standing tree and atmospheric electricity. *Comptes Rendus Geosciences*, v.342, N 2, p.95-99, 2010.

**Liperovsky V.A., Pokhotelov O.A., Meister C.-V., Liperovskaya E.V.** Physical models of coupling in the lithosphere-atmosphere-ionosphere system before earthquakes. *Geomagnetism and Aeronomy*, v.48 N 6, p.795-806, doi:10.1134/S0016793208060133, 2008.

**Liu J.Y., Chuo Y.J., Shan S.J., Tsai Y.B., Chen Y.I., Pulinets S.A., Yu S.B.** Pre-earthquake ionospheric anomalies registered by continuous GPS TEC measurements. *Ann. Geophys.*, v.22, N 5, p.1585-1593, doi:10.5194/angeo-22-1585-2004, 2004.

**Liu J.Y., Chen Y.I., Chuo Y.J., Chen C.S.** A statistical investigation of preearthquake ionospheric anomaly. *J. Geophys. Res.*, v.111, p.A05304, doi:10.1029/2005JA011333, 2006.

**Liu J.Y., Le H., Chen Y.I., Chen C.H., Liu L., Wan W., Su Y.Z., Sun Y.Y., Lin C.H., Chen M.Q.** Observations and simulations of seismoionospheric GPS total electron content anomalies before the 12 January 2010 M7 Haiti earthquake. *J. Geophys. Res.*, v. 116(A04302), doi:10.1029/2010JA015704, 2011.

**Making M., Ogawa T.** Responses of atmospheric electric field and air-earth current to variations of conductivity profiles. *J. Atmos. Terr. Phys.*, v.46, N 5, p.431-445, 1984.

**Mareev E.A.** Global electric circuit research: Achievements and prospects. *Uspekhi Fizicheskih Nauk*, v.180, N 5, p.527-534, doi:10.3367/UFNr.0180.201005h.0527, 2010.

**Molchanov O.A.** Precursory effects in the subionospheric VLF signals for the Kobe earthquake. *Physics Earth Planet. Inter.*, v.105, N 3-4, p.239-248, doi:10.1016/S0031-9201(97)00095-2, 1998.

**Moore G.W.** Magnetic disturbances preceding the 1964 Alaska earthquake. *Nature*, v.203, Iss. 4944, p.508-509, doi:10.1038/203508b0, 1964.

**Namgaladze A.A., Korenkov Yu.N., Klimenko V.V., Karpov I.V., Bessarab F.S., Surotkin V.A., Glushchenko T.A., Naumova N.M.** Global model of the thermosphere-ionosphere-protonosphere system. *Pure and Applied Geophysics*, v.127, N 2/3, p.219-254, doi:10.1007/BF00879812, 1988.

**Namgaladze A.A., Korenkov Yu.N., Klimenko V.V., Karpov I.V., Surotkin V.A., Naumova N.M.** Numerical modeling of the thermosphere-ionosphere-protonosphere system. *J. Atmos. Terr. Phys.*, v.53, N 11/12, p.1113-1124, doi:10.1016/0021-9169(91)90060-K, 1991.

**Namgaladze A.A., Martynenko O.V., Volkov M.A., Namgaladze A.N., Yurik R.Yu.** High-latitude version of the global numeric model of the Earth's upper atmosphere. *Proc. of the MSTU*, v.1, N 2, p.23-84, 1998a. URL: http://goo.gl/8x9f2.

**Namgaladze A.A., Martynenko O.V., Namgaladze A.N.** Global model of the upper atmosphere with variable latitudinal integration step. *Int. J. of Geomagnetism and Aeronomy*, v.1, N 1, p.53-58, 1998b.

**Namgaladze A.A., Shagimuratov I.I., Zakharenkova I.E., Zolotov O.V., Martynenko O.V.** Possible mechanism of the TEC enhancements observed before earthquakes. *XXIV IUGG General Assembly, Perugia, Italy, 02-13 July 2007, Session JSS010*, 2007.

**Namgaladze A.A., Klimenko M.V., Klimenko V.V., Zakharenkova I.E.** Physical mechanism and mathematical modeling of earthquake ionospheric precursors registered in total electron content. *Geomagnetism and Aeronomy*, v.49, N 2, p.252-262, doi:10.1134/S0016793209020169, 2009a.

**Namgaladze A.A., Zolotov O.V., Zakharenkova I.E., Shagimuratov I.I., Martynenko O.V.** Ionospheric total electron content variations observed before earthquakes: Possible physical mechanism and modeling. *Proc. of the MSTU*, v.12, N 2, p.308-315, 2009b, ArXivID: 0905.3313 URL: http://goo.gl/A8cLx.

**Namgaladze A.A.** Physical model of earthquake ionospheric precursors (Invited). *Abstract NH24A-03 presented at 2010 Fall Meeting*, AGU, San Francisco, Calif., 13-17 Dec., 2010.

**Namgaladze A.A., Zolotov O.V.** Ionospheric effects from different seismogenic electric field sources. *XXXth URSI General Assembly and Scientific Symposium*, Istanbul, Turkey. IEEE, August. doi:10.1109/URSIGASS.2011.6051040, 2011.

**Namgaladze A.A., Zolotov O.V., Prockhorov B.E.** The TEC signatures as strong seismic event precursors. XXXth URSI General Assembly and Scientific Symposium. Istanbul, Turkey. IEEE, August. doi:10.1109/URSIGASS.2011.6051048, 2011a.

**Namgaladze A.A., Zolotov O.V., Prokhorov B.E.** Ionospheric TEC effects related to the electric field generated by external electric current flowing between faults and ionosphere. *Physics of Auroral Phenomena: Abstracts of 34 Annual Seminar (Apatity, March 1-4, 2011), Apatity*, p.50, 2011b.

**Namgaladze A.A., Ferster M., Prokhorov B.E., Zolotov O.V.** Electromagnetic drivers in the upper atmosphere: Observations and modeling. *In: "The Atmosphere and Ionosphere: Elementary Processes, Discharges and Plasmoids (Physics of Earth and Space Environments)", Bychkov V., Golubkov G., Nikitin A. (Eds.), Springer*, 300 p., 2012.

**Namgaladze A.A., Zolotov O.V.** Ionospheric effects of seismogenic disturbances of the global electric circuit seismogenic disturbances. *In: "Earthquakes: Triggers, Environmental Impact and Potential Hazards", Kostas Konstantinou (Ed.)*, NovaPub, 2012.

**Noll C.** The crustal dynamics data information system: A resource to support scientific analysis using space geodesy. *Adv. Space Res.*, v.45, N 12, p.1421-1440, 2010.

**Ondoh T.** Investigation of precursory phenomena in the ionosphere, atmosphere and groundwater before large earthquakes of M>6.5. *Advances in Space Research*, v.43, N 2, p.214-223, doi:10.1016/j.asr.2008.04.003, 2009.

**Parrot M., Mogilevsky M.** VLF emissions associated with earthquakes and observed in the ionosphere and the magnetosphere. *Physics of the Earth and Planetary Interiors*, v.57, N (1-2), p.86-99, doi: 10.1016/0031-9201(89)90218-5, 1989.

**Pulinets S.A.** Seismic activity as a source of the ionospheric variability. *Adv. Space Res.*, v.22, N 6, p.903-906, 1998.

**Pulinets S.A., Legen'ka A.D., Gaivoronskaya T.V., Depuev V.Kh.** Main phenomenological features of ionospheric precursors of strong earthquakes. *J. Atmos. Solar-Terr. Physics*, v.65, p.1337-1347, doi:10.1016/j.jastp.2003.07.011, 2003.

**Pulinets S.A., Boyarchuk K.** Ionospheric precursors of earthquakes. *Springer, Berlin, Germany*, 315 p., 2004.

**Ruzhin Yu.Ya., Oraevsky V.N., Shagimuratov I.I., Sinelnikov V.M.** Ionospheric precursors of earthquakes revealed from GPS data and their connection with "sea-land" boundary. *Proc. 16th Wroclaw EMC Symposium*, p.723-726, 2002.

**Sorokin V.M., Chmyrev V.M., Yaschenko A.K.** Theoretical model of DC electric field formation in the ionosphere stimulated by seismic activity. *J. Atmos. Solar-Terr. Phys.*, v.67, p.1259-1268, 2005.

**Sorokin V.M., Yaschenko A.K., Hayakawa M.** Formation mechanism of the lower-ionosphere disturbances by the atmosphere electric current over a seismic region. *J. Atmos. Solar-Terr. Phys.*, v.68, p.1260-1268, 2006.

**Sorokin V.M., Yaschenko A.K., Hayakawa M.** A perturbation of DC electric field caused by light ion adhesion to aerosols during the growth in seismic-related atmospheric radioactivity. *Nat. Haz. Earth System Sci.*, v.7, p.155-163, 2007.

**Uyeda S., Nagao T., Kamogawa M.** Short-term earthquake prediction: Current status of seismo-electromagnetics. *Tectonophysics*, v.470, N 3-4, p.205-213, doi:10.1016/j.tecto.2008.07.019, 2009.

**Wessel P., Smith W.H.F.** New, improved version of Generic Mapping Tools released. *EOS trans.*, v.79, p.579, 1998.

**Xu T., Chen Z., Li Ch., Wu J., Hu Ya., Wu Z.** GPS total electron content and surface latent heat flux variations before the 11 March 2011 M9.0 Sendai earthquake. *Adv. Space Res.*, v.48, p.1311-1317, doi:10.1016/j.asr.2011.06.024, 2011a.

**Xu T., Hu Ya., Wu J., Wu Z., Li Ch., Xu, Suo Y.** Anomalous enhancement of electric field derived from ionosonde data before the great Wenchuan earthquake. *Adv. Space Res.*, v.47, N 6, p.1001-1005, doi:10.1016/j.asr.2010.11.006, 2011b.

**Yokoyama T., Yamamoto M., Pfaff R.F., Fukao S., Iwagami N.** SEEK-2 campaign measurement of the electric field in the E-region and its association with the QP echoes. *Abstracts for the 112th SGEPSS Fall Meeting. Tokyo University of Electro-Communications*, p.12-13, 2002.

**Zakharenkova I.E., Krankowski A., Shagimuratov I.I.** Modification of the low-latitude ionosphere before December 26, 2004 Indonesian earthquake. *Nat. Haz. Earth System Sci.*, v.6, p.817-823, 2006.

**Zakharenkova I.E., Shagimuratov I.I., Krankowski A., Lagovsky A.F.** Precursory phenomena observed in the total electron content measurements before great Hokkaido earthquake of September 25, 2003 (M=8.3). *Studia Geophysica et Geodaetica*, v.51, N 2, p.267-278, 2007.

**Zolotov O.V., Prokhorov B.E, Namgaladze A.A., Martynenko O.V.** Variations of the ionosphere Total Electron Content before earthquakes. *Russian Journal of Physical Chemistry B*, v.5, N 3, p.435-438, doi:10.1134/S1990793111030146, 2011.

**Zolotov O.V., Namgaladze A.A., Zakharenkova I.E., Martynenko O.V., Shagimuratov I.I.** Physical interpretation and mathematical simulation of ionospheric precursors of earthquakes at midlatitudes. *Geomagnetism and Aeronomy*, v.52, N 3, p.390-397, doi:10.1134/S0016793212030152, 2012.